# Observing Nucleation and Crystallization of Rocksalt LiF from Molten State through Molecular Dynamics Simulations with Refined Machine-Learned Force Field

Boyuan Xu[a], Liyi Bai[a,*], Shenzhen Xu[b,c], Qisheng Wu[a,*]

[a]Suzhou Laboratory, Suzhou, Jiangsu 215123, People's Republic of China
[b]School of Materials Science and Engineering, Peking University, Beijing 100871, People's Republic of China
[c]AI for Science Institute, Beijing 100084, People's Republic of China

*Corresponding authors: wuqs@szlab.ac.cn; baily@szlab.ac.cn

**Abstract**

Lithium fluoride (LiF) is a critical component for stabilizing lithium metal anode and high-voltage cathodes towards the next-generation high-energy-density lithium batteries. Recent modeling study reported the formation of *wurtzite* LiF below ~550 K (*J. Am. Chem. Soc.* 2023, 145, 1327-1333), in contrast to experimental observation of *rocksalt* LiF under ambient conditions. To address this discrepancy, we employ molecular dynamics (MD) simulations with a refined machine-learned force field (MLFF), and demonstrate the nucleation and crystallization of *rocksalt* LiF from the molten phase at temperatures below ~800 K. The *rocksalt* phase remains stable in LiF nanoparticles. Complementary density functional theory (DFT) calculations show that dispersion interactions are essential for correctly predicting the thermodynamic stability of *rocksalt* LiF over the *wurtzite* phase on top of the commonly used PBE functional. Furthermore, we show that inclusion of virial stresses—alongside energies and forces—in the training of MLFFs is crucial for capturing phase nucleation and crystallization of *rocksalt* LiF under the isothermal-isobaric ensemble. These findings underscore the critical role of dispersion interactions in atomistic simulations of battery materials, where such effects are often non-negligible, and highlight the necessity of incorporating virial stresses during the training of MLFF to enable accurate modeling of solid-state systems.

# 1. Introduction

As global energy demand escalates and the urgency to mitigate fossil fuel reliance intensifies, the development of advanced sustainable technologies, especially those of enhanced energy density, has become critical to address the fast-approaching physicochemical limitations of conventional lithium (Li)-ion batteries.[1-3] Innovations such as Li metal anodes, high-voltage cathode materials, and next-generation electrolytes—including advanced liquid electrolytes[4,5] and solid-state electrolytes[6] — have emerged as promising solutions. However, Li metal anode and high-voltage cathode face significant interfacial challenges,[7-9] such as dendrite growth and parasitic side reactions,[10-12] which degrade battery performance, accelerate capacity fade, and pose safety risks. In this context, the formation of a robust solid-state electrolyte interphase (SEI)[13], which is ionically conductive yet electronically insulating, is essential for stabilizing the Li metal anode. Simultaneously, a stable cathode-electrolyte interphase (CEI)[14] is necessary to enable the practical application of high-voltage cathodes (**Figure 1a**).

Lithium fluoride (LiF) with *rocksalt* structure (**Figure 1b**) stands out as a critical component in both SEI and CEI layers owing to its exceptional mechanical robustness, superior electrical insulation properties, and wide electrochemical stability window, which collectively contribute to the enhanced battery performance and prolonged cycle life.[15,16] Nevertheless, the intrinsically low ionic conductivity of LiF remains a key limitation, as Li ion diffusion through the inorganic inner layer has been identified as the rate-limiting step in interfacial charge transfer.[17,18] To mitigate these challenges, LiF must be synergistically combined with secondary phases (e.g., $Li_2O$ and $Li_2CO_3$) to construct hybrid interfaces that balance ionic transport and mechanical integrity.[19] Experimental characterizations of SEI and CEI, however, remain extremely challenging due to their inherent structural complexity and dynamic evolution,[14,20] which underscores the critical necessity of computational modeling to unravel their atomic structures, interfacial interactions, and ion transport mechanisms.[21] While density functional theory (DFT) and *ab initio* molecular dynamics (AIMD) methods were employed in early studies to probe the electronic properties and diffusion mechanisms of SEI systems,[22-24] they were restricted to small systems and short simulation times due to high computational cost, and thus more advanced computational approaches are required to simulate larger-scale, disordered structures over extended timescales for a deeper understanding of the structures evolution of SEI.

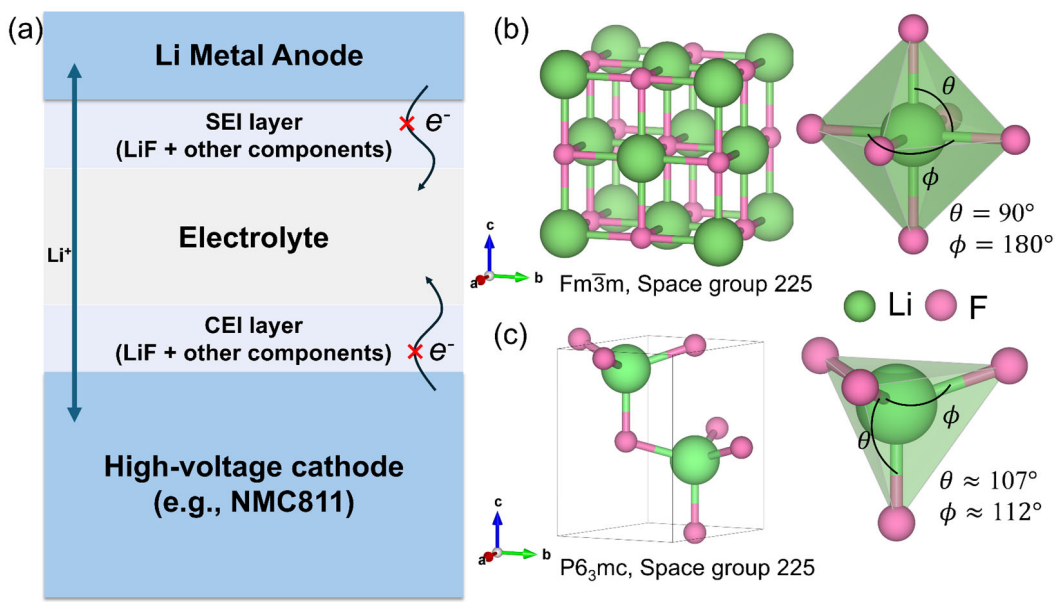

**Figure 1.** (a) Schematic for lithium batteries highlighting the LiF-containing SEI and CEI layers, where LiF can protect both Li metal anode and high-voltage cathode. (b) *Rocksalt* phase of LiF featuring octahedral coordination between Li and F. (c) *Wurtzite* phase of LiF featuring tetrahedral coordination between Li and F. The representative angles between neighboring Li-F bonds in both phases are marked. The Li and F atoms are colored with green and pink, respectively. CsCl-type and zinc blende LiF structures are shown in **Figure S1**.

Machine-learned force fields (MLFFs) leverage neural networks to construct potential energy surfaces directly from quantum-level training data, enabling the prediction of energies and forces without relying on predefined bonding rules.[25] This enables MD simulations of a system with tens of thousands of atoms over nanoseconds and beyond with quantum-level accuracy.[26] MLFFs have proven invaluable in uncovering the ion-transport mechanisms and structural evolution in solid-state electrolytes,[27,28] electrode materials[29,30] and interfacial systems[31,32] of lithium batteries. Despite their advantages, MLFFs carry an inherent limitation rooted in their foundational design: their predictive accuracy hinges critically on the data accuracy, coverage and diversity of DFT training datasets. Therefore, rigorous validation against both DFT calculations and experimental observations—spanning both structural and dynamic properties—is essential to ensure the reliability of the MLFFs.[33-35]

A recent MD study[31] with MLFFs investigated atomic structures and lithium-ion transport behavior in amorphous inorganic SEI components composed of LiF and $Li_2CO_3$, in which it was found that the presence of $Li_2CO_3$ suppresses the nucleation and crystallization of LiF, thereby promoting the ionic conductivity of SEI. In absence of $Li_2CO_3$, LiF was found to crystallize into the *wurtzite* phase, characterized by Li-F tetrahedra, at temperatures below ~550 K. While this study offers valuable insights into the impact of the local environmental on Li ion diffusion in the major SEI components, one must notice that the computational observation of *wurtzite* LiF ($P6_3mc$, Space Group 186, **Figure 1c**) stands against the experimental observation of *rocksalt* LiF ($Fm\bar{3}m$, Space Group 225, **Figure 1b**)[36] under ambient condition. The *rocksalt* LiF remains stable even under extreme pressures (up to 100 GPa) and temperatures approaching its melting point.[37,38]

In this work, we have conducted comprehensive DFT calculations to identify the thermodynamically stable phases of LiF, alongside molecular dynamics (MD) simulations with a refined MLFF to observe the nucleation and crystallization of *rocksalt* LiF from its molten state. We found that, when using the Perdew-Burke-Ernzerhof (PBE) functional,[39] the inclusion of dispersion interaction (e.g., the DFT-D3 method of Grimme[40]) is critical in accurately predicting the energetic preference of *rocksalt* LiF over *wurtzite* LiF. This finding is further supported by calculations using the regularized version of the strongly constrained and appropriately normed functional ($r^2$SCAN)[41]. These results suggest that the dispersion interaction must be explicitly considered during the training of MLFFs for accurate phase stability predictions. Additionally, we demonstrate that the incorporation of virial stress during force field training is essential to observe the nucleation and crystallization processes of *rocksalt* LiF in MD simulations. Leveraging advancements in our refined MLFF, we achieved high-fidelity MD trajectories over tens of nanoseconds, enabling direct observation of the nucleation of *rocksalt* LiF and its subsequent crystallization from the molten state. With these findings, we highly recommend that one carefully validates the choice of DFT method, the force field training process and the MLFF-informed molecular dynamics simulations.

## 2. Methods

**DFT Calculation Setup.** DFT calculations were performed using the Projector Augmented Wave (PAW) method[42] implemented in the Vienna Ab initio Simulation Package (VASP). For benchmarking purposes, the PBE[39], the local density approximation (LDA) of Slater exchange[43] with Perdew-Zunger parametrization of Ceperley-Alder Monte Carlo correlation data[44,45], and the $r^2$SCAN methods[41] were used for the DFT exchange correlation functional. The PBE-D3 method with Becke-Johnson damping function[40] and r²SCAN+rVV10 (b=11.95)[46] were selected for van der Waals correction. The plane-wave cutoff energy was set to 800 eV with 0.05 Å$^{-1}$ Gamma-centered k-point meshes for static calculations and geometry optimizations of 8-atom cell of *rocksalt* LiF (Fm$\bar{3}$m, Space Group 225) and 4-atom cell of wurtzite LiF (P6$_3$mc, Space Group 186), respectively. The convergence criteria for energy and force were set to $1 \times 10^{-6}$ eV and 0.001 eV Å$^{-1}$, respectively. For static calculations of AIMD snapshots, the GGA-PBE-D3 and *r²*SCAN were used with cutoff energy of 600 eV, k-point mesh of 0.5 Å$^{-1}$, and energy convergence criterion of $1 \times 10^{-6}$ eV.

**AIMD Simulation Setup**. AIMD simulations with the constant number, volume, and temperature (NVT) ensemble were conducted with PBE-D3 and *r²*SCAN for both *rocksalt* LiF ($2 \times 2 \times 2$ and $3 \times 3 \times 3$ supercell) and *wurtzite* LiF ($2 \times 2 \times 2$ and $3 \times 3 \times 3$ supercell). The plane-wave cutoff energy was set to 500 eV with 0.5 Å$^{-1}$ k-point meshes. The timestep was set to 1 fs, and the convergence criteria for electronic energy and ionic energy self-consistent loops were set to $1 \times 10^{-6}$ eV and $1 \times 10^{-5}$ eV, respectively. For simulations with PBE-D3, a total of 216 snapshot structures were extracted from 2 ps AIMD trajectories (the first 200 fs trajectories were discarded) under 300 K, 600 K and 1200 K. For simulations with *r²*SCAN, a total of 574 structures were extracted from AIMD trajectories (10 ps for $2 \times 2 \times 2$ and 2.8 ps for $3 \times 3 \times 3$ *rocksalt* supercell; 7.5 ps for $2 \times 2 \times 2$ and 5.3 ps for $3 \times 3 \times 3$ *wurtzite* supercell; the first 1 ps trajectories were discarded) at 300 K, 600 K, 900 K and 1200 K.

**Machine Learning Molecular Dynamics (MLMD) with training dataset from the literature**

The complete force field training and MLMD simulations are illustrated in **Figure S2**. The training dataset was obtained from previous work[31] and the model was trained for $1\times 10^6$ steps with DeePMD-kit[47,48]. The embedding network consisted of three layers with 25, 50, and 100 nodes, respectively, while the fitting network comprised three layers with 240 nodes each. The Adam method[49] was used to minimize the loss functions with an exponential decay learning rate from $1.00\times 10^{-3}$ to $3.51\times 10^{-8}$. To construct PBE-D3 models without performing additional DFT calculations, the energies, forces and virials of the original dataset were corrected using Grimme's D3 dispersion scheme with Becke-Johnson damping, as implemented in the associated toolset [50]. By following the reported setup[31], the initial structure was chosen as a 960-atom $21.98\times 21.98\times 21.98$ Å$^3$ cubic supercell with randomly packed 480 Li-F pairs generated using Packmol.[51] The MLMD simulations were carried out through the interface of the DeePMD-kit within LAMMPS[52]. The Nosé-Hoover thermostat[53,54] was used for temperature control. Models trained without virial data were applied in the NVT ensemble, whereas those trained with virial data were used for NPT simulations. A uniform timestep of 1 fs was used throughout. All systems were initiated at 1200 K and then gradually cooled to 300 K. MLMD runs of 200 ps were conducted at high temperatures ($>600$ K), while extended 20 ns simulations were carried out at lower temperatures (< 500 K). During the intermediate temperature range (600-500 K), 5 ns simulations were used to capture the transition behavior. NPT simulations followed a similar protocol, except that the phase transition region was observed between 700 K and 900 K, during which nucleation and crystallization occurred as the Li and F stopped diffusion and presented a well-arranged lattice.

**MLMD with our in-house training dataset**

**DP-GEN Iterations.** We employed the DP-GEN workflow[55] to first train four models using DeePMD-kit with our own AIMD snapshots based on PBE-D3 or $r^2$SCAN functionals. The initial dataset also included randomly perturbed (1-5% strain) ordered *wurtzite* and *rocksalt* LiF supercells (40 structures). Each DP model was trained for $4\times10^6$ steps. The embedding network and the fitting network parameters followed the previous MLMD setup. The Adam optimizer was used to minimize the loss function, with an exponentially decaying learning rate from $1.00\times 10^{-3}$ to $3.51\times 10^{-8}$. Model deviation $\varepsilon$ for each configuration was defined as the maximum force disagreement among the four models. Configurations with deviation values falling within the range of [0.15, 0.25] eV Å$^{-1}$ were selected for DFT labeling and added to the training dataset for the next iteration. During the exploration stage, all snapshots obtained from the 40 ps MLMD NPT trajectories (spanning 300 K to 1200 K) exhibited force deviations below 0.15 eV Å$^{-1}$, indicating that the DP-GEN iterations had converged. A total of 551 and 775 configurations were collected for PBE-D3- and $r^2$SCAN-based DPGEN iterations, respectively. Subsequent MLMD NPT simulations were then carried out using these refined models, following a workflow like that used in the initial MLMD simulations. Additional temperature points were included in the molten and transition regions, while fewer points were selected after the crystallization stage to save computational cost.

**NEP-GPUMD.** The highly efficient neuroevolution potentials (NEP) approach[56,57] along with the GPUMD code[58] were used for further validation. The training dataset was derived from our previous PBE-D3 calculations, and the Ziegler-Biersack-Littmark (ZBL) screened nuclear repulsion potential[59] was incorporated during training to ensure stability at short interatomic distances. The cutoff radii for radial and angular descriptor parts were set to 6 Å and 5 Å, respectively. For radial (angular) descriptor components, we used 6 (5) radial functions constructed from a linear combination of 10 (8) basis functions. A single hidden layer containing 50 neurons was employed, and the ZBL cutoff was set to 1.4 Å (close to the F-F bond length of 1.43 Å) to avoid unphysical atomic overlap. The NEP potential was trained using the entire training set for a total of 400,000 optimization iterations. A final root mean square errors (RMSEs) for energies, forces and virials were 0.94 meV/atom, 27.8 meV/Å and 8.7 meV/atom, respectively, where RMSE is defined as

$$RMSE = \sqrt{\frac{1}{N}\sum_{i=1}^{N}(Y_{ML} - Y_{DFT})^2}$$

where Y stands for the energies, forces or virials values.

MLMD NPT simulations were then performed using both 960-atom and 7680-atom (2 × 2 × 2 supercell of 960-atom) cells via the GPUMD interface, following the same protocols as the MLMD NPT simulations with DP models. The simulations utilized the Stochastic Cell Rescaling (SCR) barostat[60] in conjunction with the Bussi-Donadio-Parrinello (BDP) thermostat.[61] The linear and angular momenta were set to zero every 10 steps to prevent system drift. To assess the equilibration behavior of large-scale systems in the transition region, we extracted initial configurations at 800 K, 825 K, and 850 K and conducted extended NPT simulations for 50 ns. Additionally, to investigate the crystallization behavior of LiF in vacuum, molten LiF clusters containing 960, 480, 120, 60, and 30 atoms were placed in enlarged boxes with ~20 Å spacing between periodic images. These clusters underwent identical NVT cooling processes to explore size-dependent phase evolution.

Compared to the Nosé-Hoover method used with DeePMD in LAMMPS, the SCR barostat with the BDP thermostat embedded in NEP-GPUMD produced significantly less anisotropic deformation of the simulation cell. To ensure that observed differences in transition temperatures and lattice behavior were due to thermostat/barostat selection rather than differences in the force fields, we conducted a control simulation: the same 960-atom NPT run was executed using NEP within GPUMD but applying the Nosé-Hoover thermostat. The simulation reproduced the anisotropic cell elongation (**Figure S3**) and a similar transition temperature (900 K), confirming that the cell deformation was a thermostat effect. A comparative computational efficiency analysis **(Figure S4)** demonstrated that NEP-GPUMD achieved 1.6 to 5 times faster simulation speed on a single RTX 3080 GPU compared to the default DeePMD running on an A800 GPU. This highlights the excellent scalability and computational efficiency of NEP-GPUMD for long-time simulations of large systems.

## 3. Results and discussions

The LiF crystallizes in the *rocksalt* structure with $F\overline{m}3m$ space group (No.225, **Figure 1b**) and its lattice parameter is 4.026 Å.[36] In this structure, $Li^+$ and $F^-$ ions occupy alternating octahedral sites, forming the NaCl-type arrangement stabilized by strong ionic bonding. LiF has a high melting point (~1118 K)[37] and a low thermal expansion coefficient[38]. The fully relaxed atomic structures and the energies were calculated with different DFT methods for *rocksalt* LiF (**Figure 2** and **Table S1**). The PBE functional slightly overestimates the lattice parameters, while the local density approximation (LDA), PBE-D3 and $r^2$SCAN functionals slightly underestimate the lattice parameters compared to experimental values. The two characteristic angles in the Li-F octahedron of *rocksalt* LiF ($\phi \approx 180°$ and $\theta \approx 90°$) are illustrated in **Figure 1b**.

In addition to *rocksalt* phase, three hypothetical polymorphs of LiF have been proposed: the *wurtzite* structure (**Figure 1c**), the body-centered cubic (bcc) CsCl-type structure (**Figure S1a**) and the cubic zinc blende structure (**Figure S1b**). However, these polymorphs are yet to be observed experimentally, even under extreme conditions of pressure (~100 GPa) and temperature near the melting point.[37,38] The DFT-calculated relative energy differences between the four phases of LiF were shown in **Figure 2**. Notably, the PBE functional incorrectly predicts the *wurtzite* phase to be more stable than the *rocksalt* phase. In contrast, the LDA, $r^2$SCAN and $r^2$SCAN+rVV10 functionals correctly determine the energy preference of *rocksalt* phase. As a meta-GGA functional, $r^2$SCAN holds good transferable accuracy[41] and thus has been treated as reliable reference for predicting the energy order between different polymorphs. Interestingly, the PBE-D3 method—which includes empirical van der Waals corrections—aligns with the $r^2$SCAN and $r^2$SCAN+rVV10 results, underscoring the critical role of dispersion interactions in accurately predicting polymorph energy ordering on top of the PBE functional.[62] The *wurtzite* LiF features two typical bonding angles of $\theta \approx 107°$ and $\phi \approx 112°$ as marked in **Figure 1c**. The bcc CsCl-type LiF exhibits significantly higher energy (> 0.5 eV/f.u.) than the *rocksalt* phase independent of the choice of DFT methods, which is consistent with previous reports.[63] Although the zinc blende phase has a similar energy to the *wurtzite* structure, it was not observed in our molecular dynamics (MD) simulations.

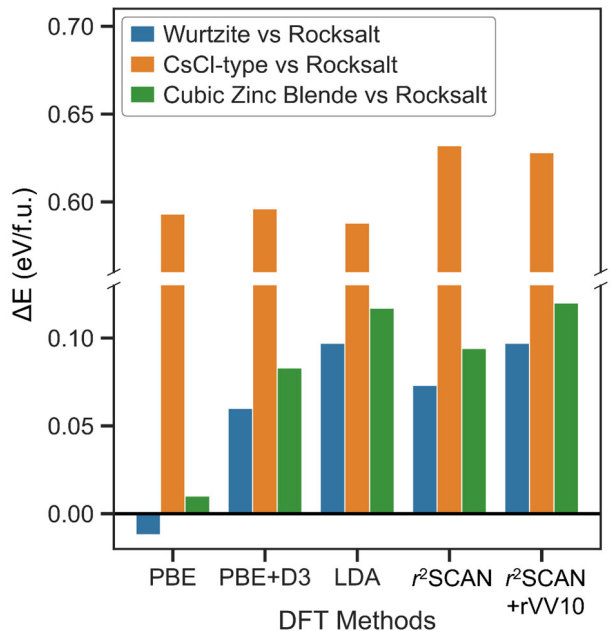

**Figure 2**. The relative energies for *wurtzite*, CsCl-type, and cubic zinc blende phases of LiF were calculated with different DFT methods, where the *rocksalt* phase was set as the reference. The f.u. stands for formula unit. The Li_sv_pot and F pseudopotentials were used in this work, and we have verified that the same energy ordering was captured when Li_pot pseudopotential was used.

To identify the key factors that impact MLMD predictions of local bonding environment in LiF, we trained four DP models based on the published LiF database.[31] These models include: 1) The model trained on energies and forces with PBE, namely **M1-PBE**; 2) The model trained on energies, forces, and virials, namely **M2-PBE-virial**; 3) The model trained on energies and forces with PBE-D3, namely **M3-PBE-D3**; 4) The model trained on energies, forces and virials with PBE-D3, namely **M4-PBE-D3-virial**. All the four models were rigorously validated against the training dataset, demonstrating high fidelity. The root-mean-square errors (RMSEs) for energies range from 0.1 to 0.2 meV/atom, those for forces from 8.0 to 11.9 meV/Å and those for virial per atom around 1.1 meV. Comparative results between DP- and DFT-calculated for energies, forces, and virials are presented in **Figure S5** and **S6**. In addition, we constructed our own dataset with energies, forces, and virials from DFT calculations of perturbed structures and AIMD simulations by PBE-D3 for independent validation, and the trained model was named as **M5-PBE-D3-virial** (see details in Methods section and the following paragraphs).

To investigate the influence of these models on local structural and dynamic properties of LiF, we performed quantitative analyses of Li-F coordination numbers and lithium diffusion coefficients under various temperature conditions. The cutoff radius of 2.5 Å was set to determine the number of Li-F neighbors, based on the radial distribution function (RDF) plots (**Figure S7**), and F-Li-F bond angles were analyzed for geometric classification (**Table S2**). For *wurtzite* LiF (**Figure 1c**), a tetrahedral structure was identified when a central Li atom was bonded to four F atoms, and all F-Li-F $\phi$ and $\theta$ bond angles fell within the range of 89.5°−129.5°, consistent with the criteria used by Hu et al.[31]. Otherwise, it was considered as Li-4F (non-tetra). For *rocksalt* LiF (**Figure 1b**), a slightly broader angle tolerance of ±20° was adopted: an octahedron was defined if the central Li atom was coordinated to six F atoms, with three of the largest F-Li-F $\phi$ angles falling within 140°−180°, and the remaining $\theta$ angles in the 70°−110° range. For each MD trajectory, the last 400 of 1000 frames were selected for coordination analysis. The diffusion coefficient $D$ was calculated using the Einstein relation:

$$D = \frac{1}{2dt} \langle [r(t)]^2 \rangle \tag{1}$$

In the case of three-dimensional diffusion ($d = 3$), the mean square displacement (MSD)[64] is given by:

$$\langle [r(t)]^2 \rangle = \frac{1}{N} \sum_{i=1}^{N} \langle [r_i(t+t_0)]^2 - [r_i(t_0)]^2 \rangle \tag{2}$$

where $r_i(t)$ is the displacement of the *i*-th Li ion at time *t*, the average $\langle ... \rangle$ is applied with respect to $t_0$ (starting time), and *N* is the number of Li atoms in the system. Diffusion coefficients were averaged every 10 frames up to 50% of the trajectory (500 frames) and reported along with standard deviations. It is worth noting that the effect of lattice volume change in NPT simulations was not removed from the MSD, which may influence the diffusion behavior, particularly during phase transitions. Decoupling these contributions is a non-trivial task, as discussed in a recent study.[65] Since precise diffusivity is not the primary focus of this work, we did not remove the effect of lattice parameter change from MSD calculations for NPT simulations.

All the MD simulations with the aforementioned MLFF models exhibited distinct signatures of LiF nucleation and crystallization, accompanied by a pronounced decrease in lithium-ion diffusion coefficients with decreasing temperature (**Figure 3** and **Figure S8**). In particular, simulations performed using the **M1-PBE** model (**Figure 3a** and **Figure S8a**) revealed a marked increase in the formation of Li-F tetrahedra around 600 K, with the corresponding diffusion behavior in good agreement with previous findings[31]. Detailed structural analyses further demonstrated that ~60% of tetrahedrons evolved from non-tetrahedral Li-4F units and ~25% from Li-5F units, indicating minimal reorganization of lithium's local coordination environments during the transition.

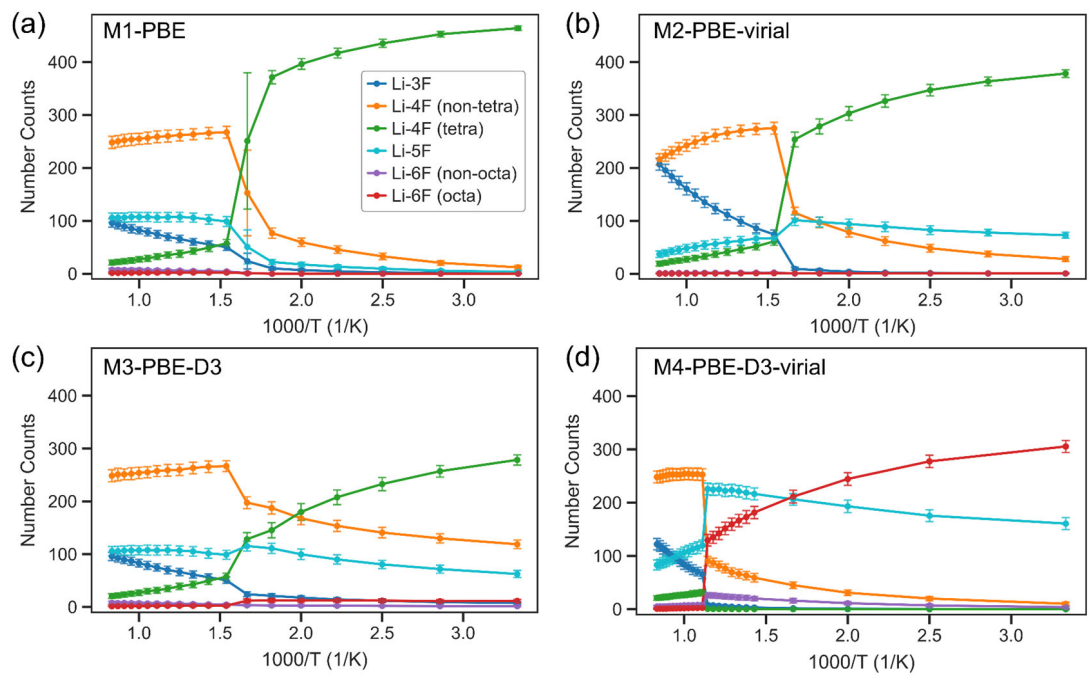

**Figure 3.** Coordination statistics for 960-atom LiF supercell from MLMD trajectories by models trained from reported dataset using (a) energy and force data for NVT simulation, (b) energy, force and virial data for NPT simulation, (c) D3 modified energy and force data for NVT simulation, (d) D3 modified energy, force and virial data for NPT simulation. As can be seen, the two NVT simulations from **M1-PBE** and **M3-PBE-D3** models and the NPT simulation with **M2-PBE-virial** model all captured a phase transition at around 600 K with the rise of LiF tetrahedrons. The NPT simulations with **M4-PBE-D3-virial** model (phase transition at 875 K) revealed an octahedron transformed from other Li-F coordination environments (Li-3F and Li-4F units).

For the **M2-PBE-virial** model (**Figure 3b** and **Figure S8b**), despite model training with virials and MD simulations under NPT ensemble, the formation of tetrahedrons remained dominant at around 600 K, with no evidence of Li-6F octahedron. Li-3F units as well as non-tetrahedral Li-4F motifs, were transformed into tetrahedral Li-4F and Li-5F units. The notable increase in Li-5F units suggests that the relaxation of lattice constraints under NPT conditions facilitates the formation of more flexible coordination environments. When dispersion corrections were introduced in the **M3-PBE-D3** model (**Figure 3c** and **Figure S8c**), the NVT simulations revealed a trend similar to that of **M2-PBE-virial** both in phase transition temperature and coordination number evolution. However, the total number of tetrahedrons was approximately 100 fewer, which was attributed to the higher retention of non-tetrahedral Li-4F units in the NVT simulation with **M3-PBE-D3** as compared to the NPT simulation with **M2-PBE-virial**. Interestingly, a small amount (~10) of octahedral Li-6F units emerged, although their population did not increase further during the simulation. It is worth noting that although the PBE-D3 functional correctly reproduces the energetic ordering between *rocksalt* and *wurtzite* LiF (**Figure 2**), the NVT ensemble imposes volumetric constraints that hinder the transformation of loosely packed Li-4F tetrahedra into the denser Li-6F octahedra characteristic of the *rocksalt* phase ($\rho_{\text{wurtzite}}/\rho_{\text{rocksalt}} \approx 80.29\%$). Under NVT simulations, this transformation only occurred in rare instances where significant local voids (~4.66% volume) formed during crystallization at 600 K (**Figure S9**).

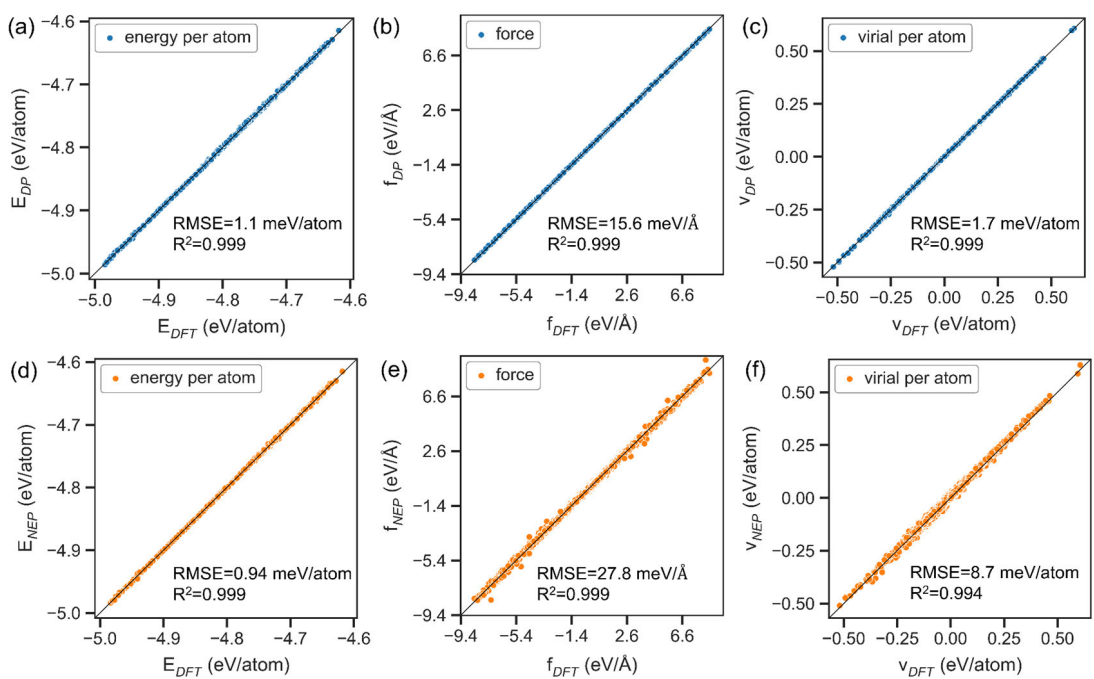

**Figure 4.** Comparisons of models trained with PBE-D3 method: (a) energy comparison for DP *vs* DFT, (b) force comparison for DP *vs* DFT, (c) virial per atom comparison for DP *vs* DFT, (d) energy comparison for NEP *vs* DFT, and (e) force comparison for NEP *vs* DFT, (f) virial per atom comparison for NEP *vs* DFT.

When NPT simulations were performed with the **M4-PBE-D3-virial** model (**Figure 3d** and **Figure S8d**), two notable differences were observed. First, the phase transition temperature increased to 875 K, representing a 275 K increase compared to that predicted by the **M1-PBE** model and much closer agreement with the experimentally measured melting point of LiF (~1118 K)[37]. The discrepancy between simulated and experimental melting or crystallization temperatures has been widely discussed, and is commonly attributed to factors such as finite-size effects and limited simulation timescales inherent to atomistic simulation,[66] the presence of impurities in experimental samples, as well as simulation methodologies and force field models employed.[67] Second, a significant reduction in Li-4F units was observed, accompanied by a pronounced increase in Li-5F and Li-6F units, indicating an overall densification of the system.

In addition to the models based on the published dataset (**M1**-**M4**), we trained a new model, **M5-PBE-D3-virial**, using our own dataset from DFT calculations of perturbed structures and AIMD simulations (see Methods and Figure 4a-c), and conducted NPT simulations with this model.[31] As shown in **Figure S10a,b**, both the coordination number evolution and the phase transition temperature (875 K, with vacuum; see **Figure S11**) closely matched those obtained using the **M4-PBE-D3-virial** model. A comparison of system density between the **M5-PBE-D3-virial** and the **M2-PBE-virial** models under NPT simulation is presented in **Figure S10c**. The simulations based on the **M5-PBE-D3-virial** model consistently yielded higher densities, underscoring the critical role of D3 dispersion corrections in accurately predicting the existence of *rocksalt* LiF. These observations highlight the importance of a well-curated dataset for force field training and an appropriate ensemble for MD simulations. Otherwise, the resulting MLMD models may lead to inaccurate or incomplete outcomes. The nucleation and crystallization snapshots at 875 K were illustrated in **Figure S10d**, revealing the gradual transformation of non-octahedral coordination units into octahedral Li-6F motifs upon cooling. To further validate our results, we conducted additional NPT simulations using a model trained with $r^2$SCAN-calculated energies, forces and virials (**M6-$r^2$SCAN-virial**, **Figure S12**). The outcomes (**Figure S13**) showed consistent trends in coordination transformation, diffusivity reduction, and phase transition temperature (~850 K). The predicted number density reached ~2.6 g/cm³, similar to that obtained with **M5-PBE-D3-virial**, confirming the robustness of the predictions. Notably, both models trained with our in-house dataset predicted a higher population of octahedral Li-6F units after the phase transition, compared to the **M4-PBE-D3-virial** model. This discrepancy is likely due to the explicit inclusion of *rocksalt* LiF and *wurtzite* LiF structures in our training dataset, which were absent in the dataset used for training the **M4-PBE-D3-virial** model.

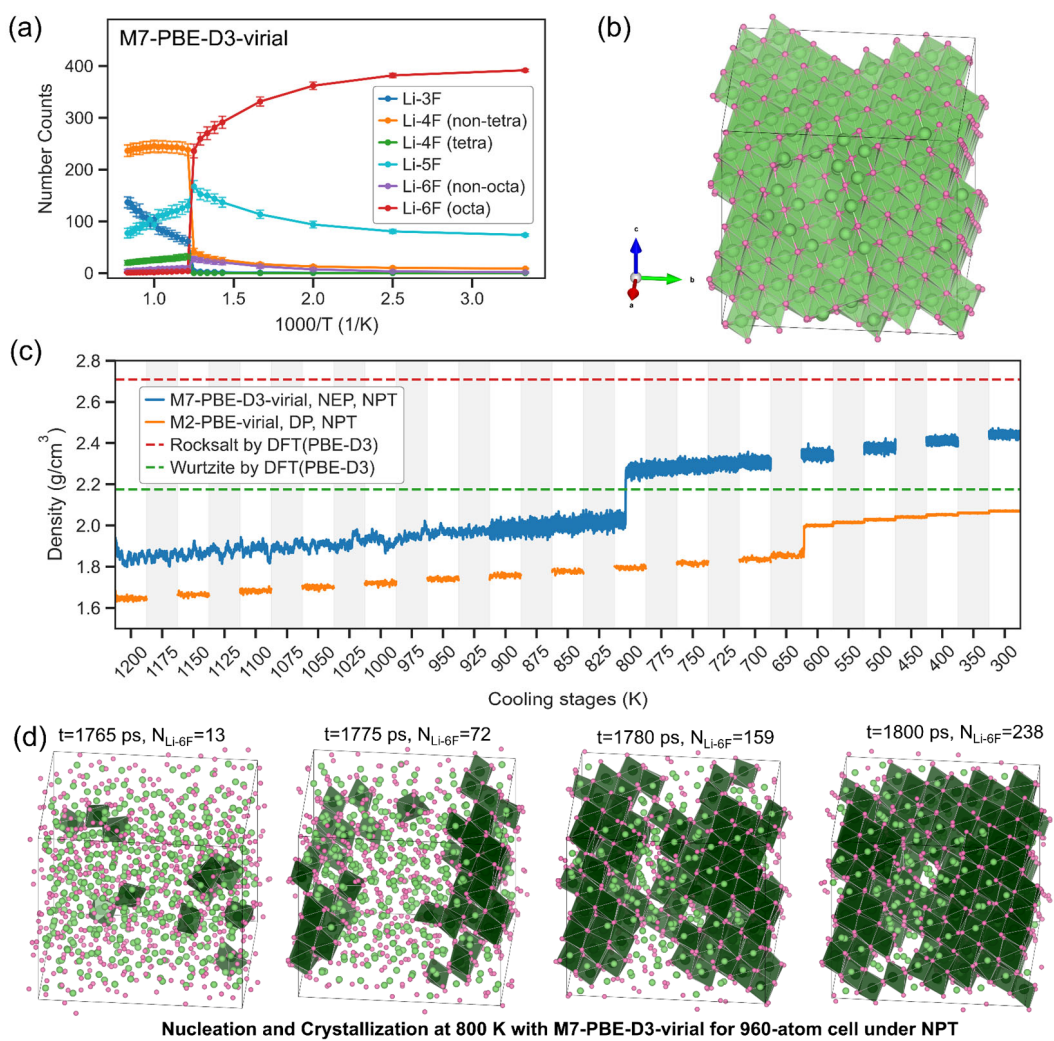

**Figure 5.** NEP-ZBL model trained with our own dataset with PBE-D3 from AIMD simulations and random perturbations of ordered *wurtzite* and *rocksalt* LiF supercells.

(a) Coordination statistics for 960-atom LiF supercell from MLMD trajectories. (b) Defected structure with obvious vacuum space (1.61%) at the last frame of 300 K simulation was plotted. (c) Comparisons of density evolution through NPT simulations with **M2-PBE-virial** and **M7-PBE-D3-virial**. The PBE-D3-calculated *wurtzite* and *rocksalt* LiF densities were shown as dashed green and red lines. (d) The nucleation and crystallization process captured by **M7-PBE-D3-virial** NPT simulation at 800 K. The Li-6F units were highlighted in dark green and their numbers were counted and listed.

The solid-state nudged elastic band (SSNEB) calculations were performed using the PBE-D3 functional to evaluate the energy barrier between the *rocksalt* and *wurtzite* phases of LiF. Potential transition pathways were selected based on previously reported mechanisms,[68] and the one with the lowest barrier energy was identified and shown in **Figure S14**. The resulting barrier was 0.16 eV per formula unit, corresponding to an estimated thermal energy ($k_B T$) of ~1850 K, which significantly exceeds the melting point of LiF (~1118 K),[37] consistent with the experimental observation that LiF stabilizes in the *rocksalt* phase under ambient conditions.

To assess the model-independence of our results, we trained a NEP-ZBL MLFF (**M7-PBE-D3-virial**) using the same dataset employed for the DP-based **M5-PBE-D3-virial** model (see Methods and **Figure 4d-f**). NPT simulations were carried out on a 960-atom supercell using an identical cooling protocol, with results summarized in **Figure 5**. The formation of the *rocksalt* phase was successfully reproduced, accompanied by increased system density and the development of vacuum regions. The observed phase transition temperature was slightly reduced to 800 K—75 K lower than that predicted by the DP model—despite the shared training dataset. This deviation likely arises from differences in the thermostat and barostat algorithms employed: the Nosé-Hoover method used in LAMMPS with DP models tends to elongate the simulation cell, whereas the SCR-BDP method implemented in GPUMD with NEP models introduces randomized rescaling factors for both cell and atomic positions under NPT conditions,[60] resulting in more physically realistic cell fluctuations. For this reason, we adopted the SCR-BDP approach for reporting MLMD results in the main text. The evolution of Li-6F coordination units during nucleation is illustrated in **Figure 5d**. Trends in Li ion diffusivity and mean squared displacement (MSD) as a function of temperature are presented in **Figure S15**, showing good agreement with the **M5-PBE-D3-virial** results. To further validate the findings, we constructed a larger simulation cell comprising 7680 atoms ($2 \times 2 \times 2$ replication of the 960-atom cell) and conducted NPT simulations. The corresponding results are shown in **Figure S16**. In this case, the phase transition occurred upon cooling to 775 K, 25 K lower than that observed in the smaller cell. Given the larger system size, structures obtained at 800 K, 825 K, and 850 K were extracted and further equilibrated independently for 50 ns to assess phase transition behavior. As expected, crystallization was observed for the 825 K and 800 K trajectories at 5.4 ns and 1.15 ns, respectively, reflecting the stochastic nature of nucleation and the necessity of extended simulation times for larger systems.

To quantitatively assess structural compactness, we performed vacuum volume analysis (details in Supplementary Information). According to our criteria, vacuum space was defined as the existence of interatomic distances exceeding 5 Å between non-bonded atoms. Visual representations and the corresponding vacuum-to-cell volume ratios are shown in **Figure S17**. All octahedron-containing structures exhibited some degree of vacuum space, whereas tetrahedron-containing structures do not, which was due to the density increase effect of forming Li-6F units. Compared to the defected NVT case, the NPT ensemble reduced but did not fully eliminate the vacuum formed during crystallization. This suggests that the suppression of phase transitions in the LiF and Li$_2$CO$_3$ mixture wound potentially not only prevent Li diffusion drops caused by LiF ordering,[31] but also minimize potential vacuum formation, which could otherwise degrade anode performance.

Finally, we examined the crystallization behavior of molten LiF clusters of various sizes in vacuum (~20 Å separation between periodic images) under the same cooling protocols to determine the preferred surface geometry of LiF. As shown in **Figure 6**, all crystalized clusters (60 to 960 atoms) exhibited *rocksalt*-type structures, with no sign of *wurtzite*-like features, confirming the structural dominance of the *rocksalt* phase. In contrast, the 30-atom cluster failed to crystalize but kept transforming from one configuration to another during the simulation even at 300 K (**Figure S18**), which can be ascribed to the fact that these configurations were close in energies as reported by previous DFT studies.[69]

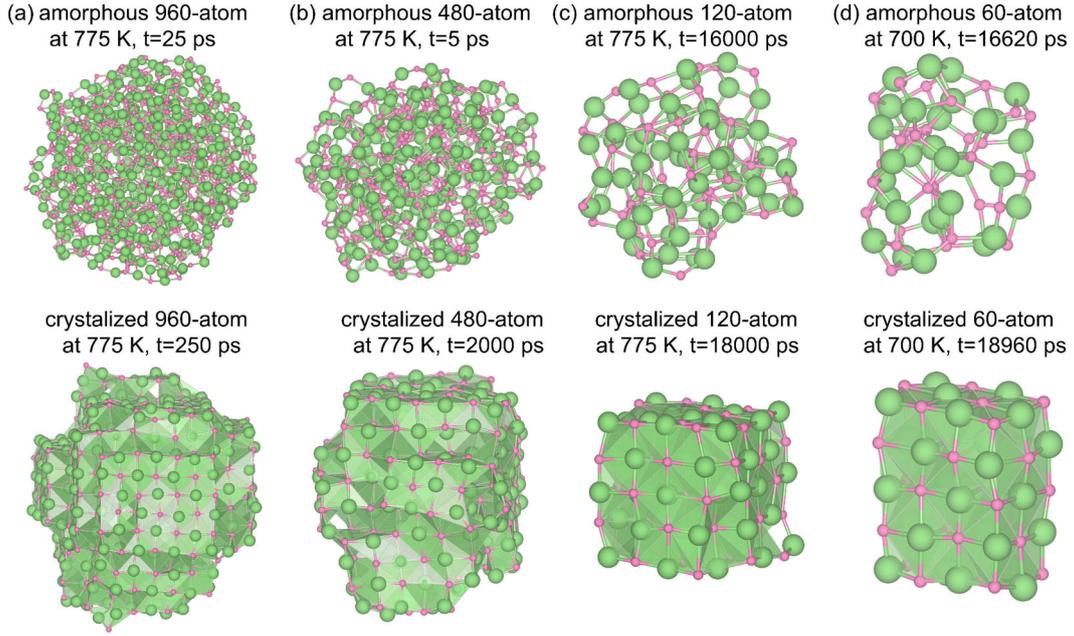

**Figure 6.** The snapshots of isolated amorphous before crystallization at phase transition temperature and crystallized LiF cluster at the same temperature with (a) 960 atoms, (b) 480 atoms, (c) 120 atoms and (d) 60 atoms.

## 4. Conclusions

In conclusion, we conducted comprehensive investigations into the atomic structure of LiF using DFT calculations and demonstrated that the inclusion of dispersion interactions on top of the PBE-D3 functional, or the use of the more advanced $r^2$SCAN functional is essential for accurately predicting the thermodynamic stability of the *rocksalt* LiF over the *wurtzite* LiF. We systematically explored the nucleation and crystallization of LiF from its molten state through MD simulations performed under both NVT and NPT ensembles, employing refined MLFFs trained on energies, forces, and virial stresses. Notably, only NPT simulations utilizing MLFFs based on PBE-D3 or $r^2$SCAN successfully reproduced the experimentally observed formation of *rocksalt*

LiF during the amorphous-to-crystalline phase transition. Complementary SSNEB calculations using PBE-D3 revealed a kinetic barrier of approximately 0.16 eV per formula unit for the transition from *rocksalt* to *wurtzite* LiF, further confirming the kinetic and thermodynamic stability of the *rocksalt* structure under ambient conditions. In addition, simulations of LiF clusters in vacuum showed that *rocksalt* symmetry is retained down to cluster sizes of approximately 60 atoms, highlighting the intrinsic structural preference of LiF. Collectively, these findings underscore the critical role of dispersion interactions and virial stress information in DFT calculations and MLMD simulations. We therefore emphasize the importance of selecting appropriate exchange-correlation functionals, rigorously training force fields, and carefully validating MD protocols when modeling battery materials and related solid-state systems.

**Data Availability:**

The training dataset for model **M5-PBE-D3-virial**, **M6-r2SCAN-virial**, and **M7-PBE-D3-virial**, are available from figshare upon the acceptance of this manuscript.

# Supplemental Material for

# Observing Nucleation and Crystallization of Rocksalt LiF from Molten State through Molecular Dynamics Simulations with Refined Machine-Learned Force Field

Boyuan Xu[a], Liyi Bai[a,*], Shenzhen Xu[b,c], Qisheng Wu[a,*]

[a] Suzhou Laboratory, Suzhou, Jiangsu 215123, People's Republic of China
[b] School of Materials Science and Engineering, Peking University, Beijing 100871, People's Republic of China;
[c] AI for Science Institute, Beijing 100084, People's Republic of China

[*]Corresponding authors: wuqs@szlab.ac.cn; baily@szlab.ac.cn

**Vacuum Volume Analysis and Visualization**

The vacuum volume analysis was conducted by following the concept of accessible free volume (AFV) analysis. Instead of the probing radius and van der Waals radius, we define the bond length cutoff to be 2.5 Å. The entire atomic structure from the POSCAR file was read, the simulation cell was discretized into a 3D grid with bins of 0.2 Å spacing (a finer bin width of 0.1 Å was tested, resulting in less than 1% variation in the calculated vacuum volume). This discretization yields a 3-dimensional array, $M_{cell}$, with binary values: 0 for non-accessible regions and 1 for accessible regions. For each atom in the structure, all grid points within a 2.5 Å radius were marked as non-accessible(i.e., assigned a value of 0 in $M_{cell}$). After iterating over all atoms, the resulting $M_{cell,ijk}$ array was exported to the voxel format using the *MidVoxIO* python module, enabling direct 3D visualization with *Magica Voxel* software. The total vacuum volume was then computed using the following formula: $V_{vacuum} = \sum_{ijk} M_{cell,ijk} \times 0.008 \text{ Å}^3$, where $0.008 \text{ Å}^3$ corresponds to the volume of each $0.2 \text{ Å} \times 0.2 \text{ Å} \times 0.2 \text{ Å}$ voxel. The structures used for vacuum volume analysis include the last frame of the following simulations: 300 K **M1-PBE** (NVT), **M2-PBE-virial** (NPT), **M3-PBE-D3** (NVT), rare case defected **M3-PBE-D3** (NVT), **M4-PBE-D3-virial** (NPT), **M5-PBE-D3-virial** (NPT), **M7-PBE-D3-virial** GPUMD simulations with 960-atom and 7680-atom NPT ensembles. Additionally, the last frame of a 50 ns 825 K **M7-PBE-D3-virial** GPUMD NPT simulation with 7680 atoms was included. All results are summarized and visualized in **Figure S17**.

**Solid-state Nudged Elastic Band (SSNEB)**

The SSNEB calculations were carried out with PBE-D3 functional, following the phase transition paths proposed by Saitta et al.[1] Both rocksalt-tetragonal-wurtzite and rocksalt-hexagonal-wurtzite paths were explored using climbing-image NEB method[2] implemented in the VASP Transition State Theory (VTST) package.[3] The transition pathway with the lowest energy barrier is reported herein. Supercells containing four atoms were employed for the calculations, and the DFT setup and convergence criteria were consistent with those used in the static calculations, as detailed in the Methods section of the main text.

**Table S1.** The lattice parameters of rocksalt, wurtzite, CsCl-type, and cubic zinc blende phases of LiF were calculated with different DFT methods.

| DFT Method | Structure | Lattice Parameter (Å) |
|---|---|---|
| PBE | Rocksalt | a=4.058 |
|  | Wurtzite | a=3.122, c=4.892 |
|  | CsCl-type | a=2.555 |
|  | Cubic Zinc Blende | a=4.359 |
| PBE+D3 | Rocksalt | a=3.992 |
|  | Wurtzite | a=3.084, c=4.810 |
|  | CsCl-type | a=2.513 |
|  | Cubic Zinc Blende | a=4.298 |
| LDA | Rocksalt | a=3.907 |
|  | Wurtzite | a=3.027, c=4.654 |
|  | CsCl-type | a=2.465 |
|  | Cubic Zinc Blende | a=4.206 |
| $r^2$SCAN | Rocksalt | a=3.985 |
|  | Wurtzite | a=3.096, c=4.721 |
|  | CsCl-type | a=2.509 |
|  | Cubic Zinc Blende | a=4.291 |
| $r^2$SCAN+rVV10 | Rocksalt | a=3.966 |
|  | Wurtzite | a=3.070, c=4.746 |
|  | CsCl-type | a=2.500 |
|  | Cubic Zinc Blende | a=4.275 |

**Table S2.** The criteria for classification of different Li-F coordination were shown here.

| | Number of F within 2.5 Å | F-Li-F angle distribution |
|---|---|---|
| Li-3F | 3 | |
| Li-4F (non-tetra) | 4 | At least one angle not in [89.5°, 129.5°] |
| Li-4F (tetra) | 4 | All angles in range [89.5°, 129.5°] |
| Li-5F | 5 | |
| Li-6F (non-octa) | 6 | Angles can't fulfill the requirement below |
| Li-6F (octa) | 6 | 3 largest angles in [140°, 180°] Other angles in [70°, 110°] |

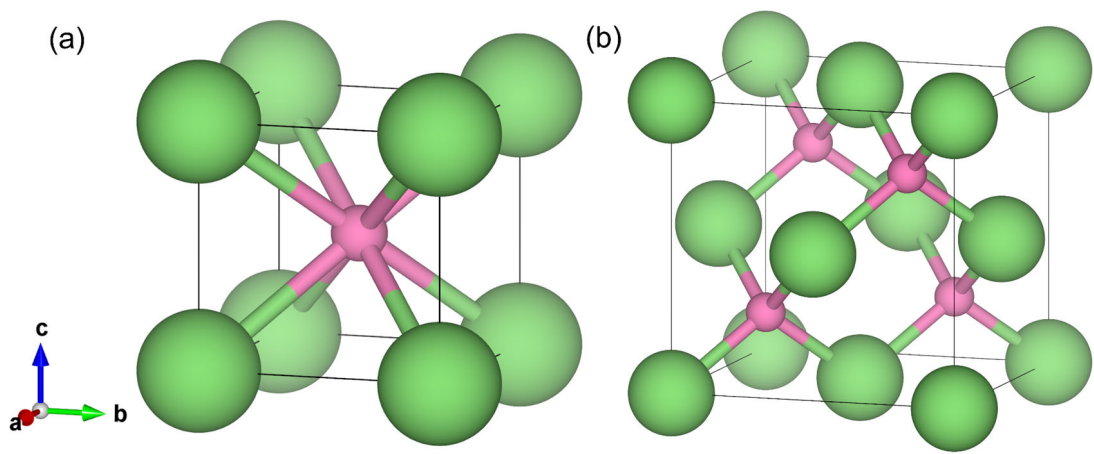

**Figure S1.** (a) CsCl-type LiF structure. (b) Cubic zinc blende LiF structure.

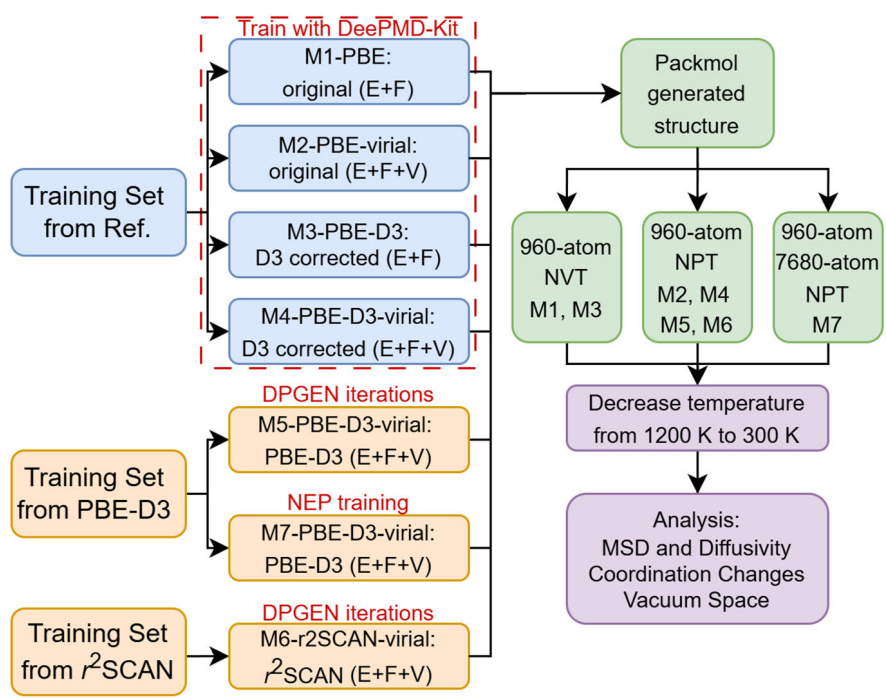

**Figure S2.** Schematic workflow for illustrating the force field training and MLMD simulations.

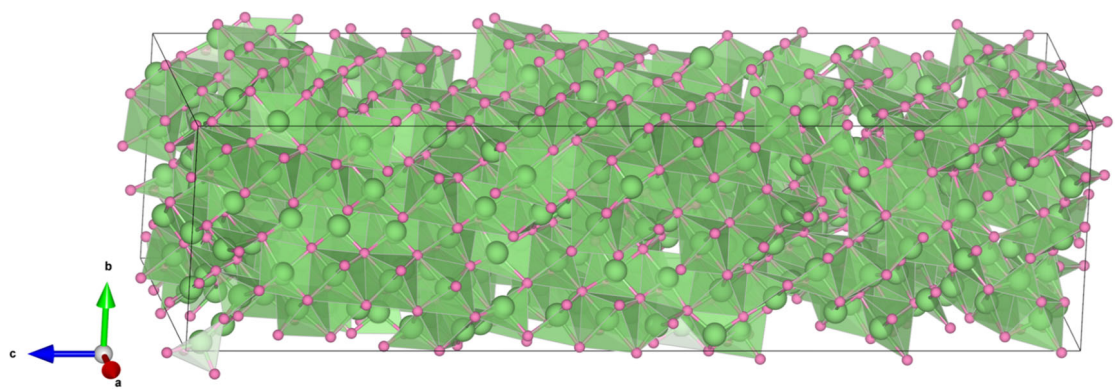

**Figure S3.** A snapshot of a specific M7 structure right after phase change at 900 K with Nosé-Hoover method of NPT simulations, showing apparent elongation along c-axis.

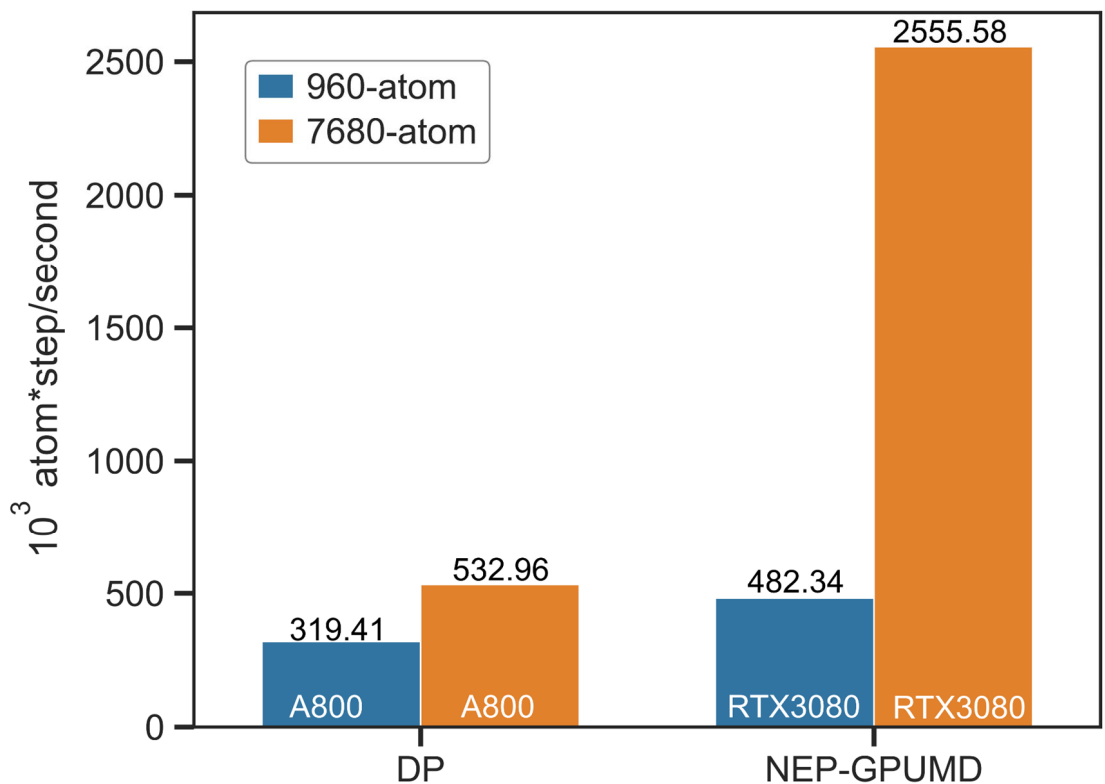

**Figure S4.** Performance comparisons between MLMD with DP and NEP-GPUMD.

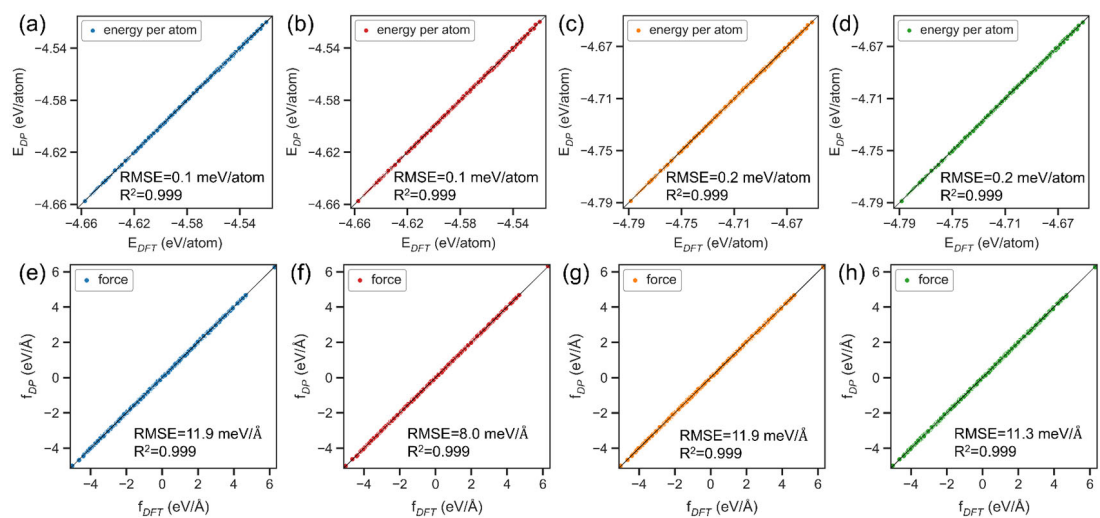

**Figure S5.** RMSE of the energies and forces by DP compared to DFT on the training dataset for different models of LiF: (a,e) **M1-PBE** (blue), (b,f) **M2-PBE-virial** (red), (c,g) **M3-PBE-D3** (orange) and (d,h) **M4-PBE-D3-virial** (green).

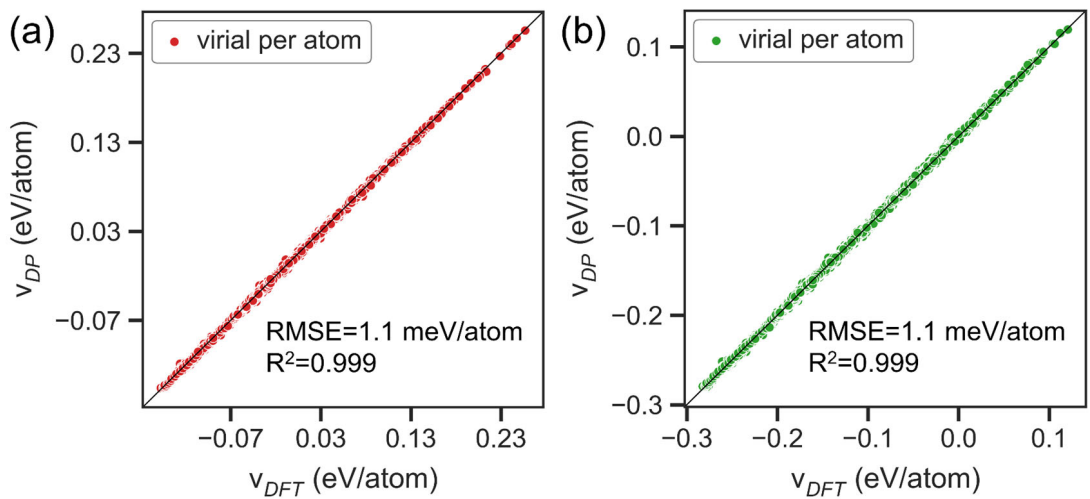

**Figure S6.** RMSE of the virial per atom by DP compared to DFT on the training dataset for different models of LiF: (a) **M2-PBE-virial** and (b) **M4-PBE-D3-virial**.

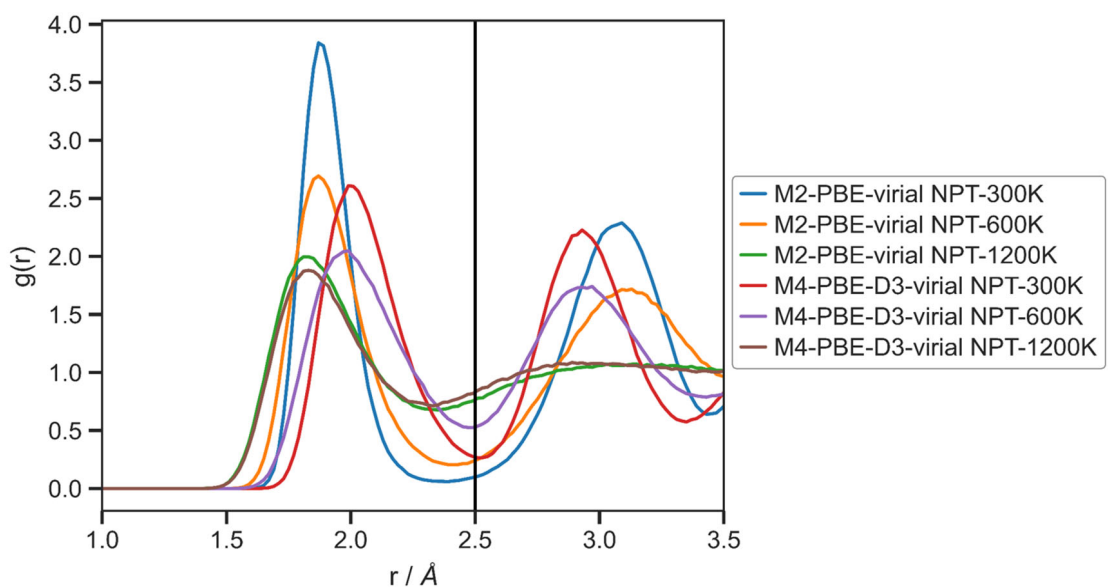

**Figure S7.** The Li radial distribution function (RDF) of **M2-PBE-virial** and **M4-PBE-D3-virial** NPT simulations at 300, 600 and 1200 K. The RDF calculations included the last 500 frames of each simulation. A vertical line at $r$=2.5 Å was chosen as the cutoff radius, demonstrating clear separation of Li's first coordination shell.

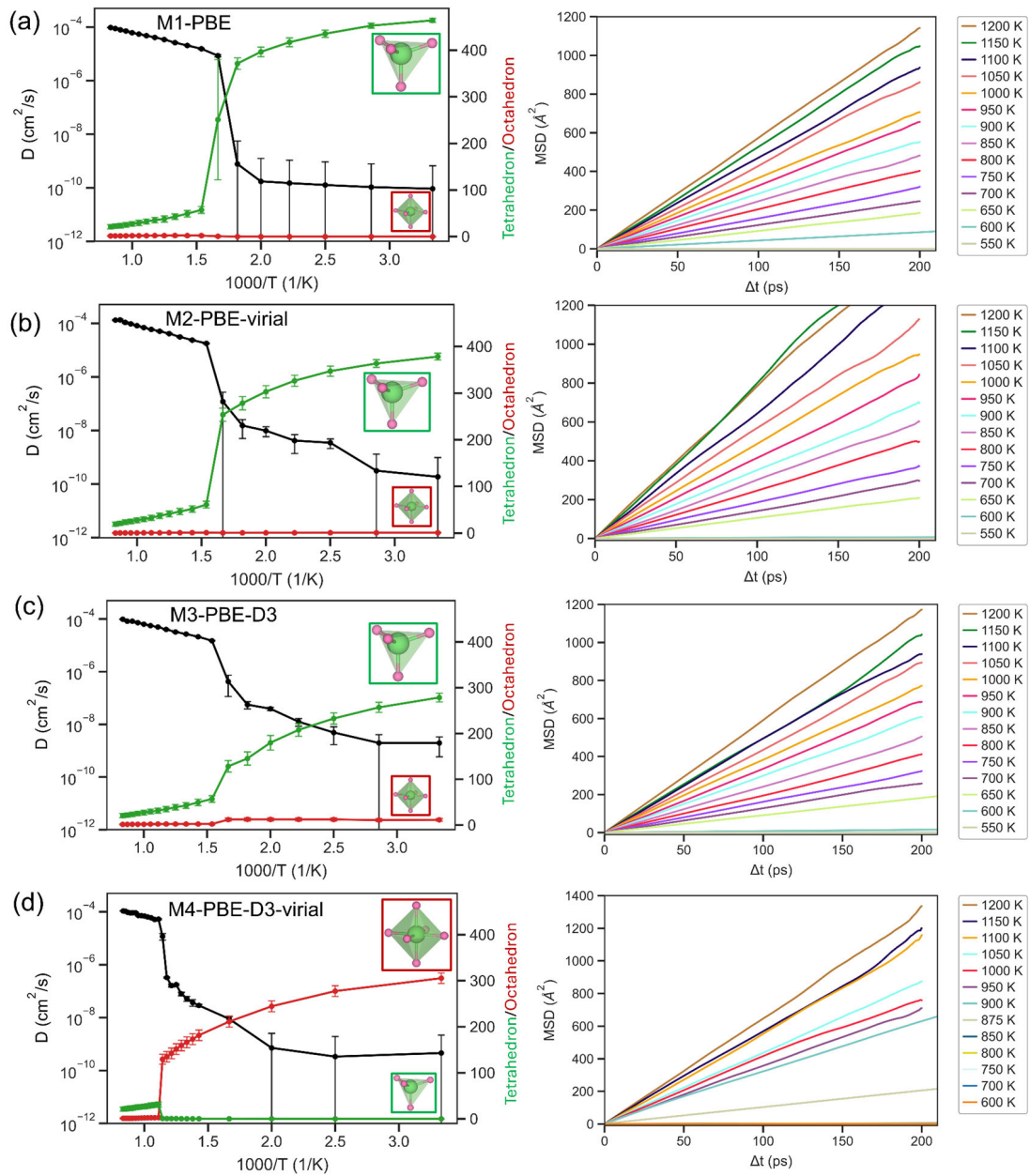

**Figure S8.** Li diffusivities (**Left**) and first 200 ps MSD (**Right**) in LiF. The defined octahedron (red) and tetrahedron (green) numbers in the LiF as a function of temperatures were plotted for (a) **M1-PBE**, (b) **M2-PBE-virial**, (c) **M3-PBE-D3** and (d) **M4-PBE-D3-virial**.

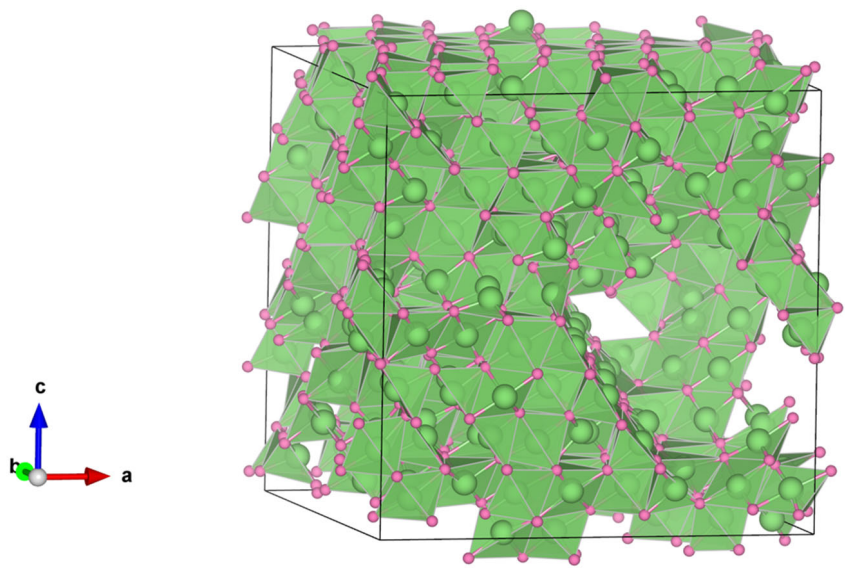

**Figure S9.** A snapshot was taken from the NVT simulation with **M3-PBE-D3** model right after phase change at 600 K, showing the existence of huge vacuum space (4.66%) along with around 100 Li-6F octahedrons at right side.

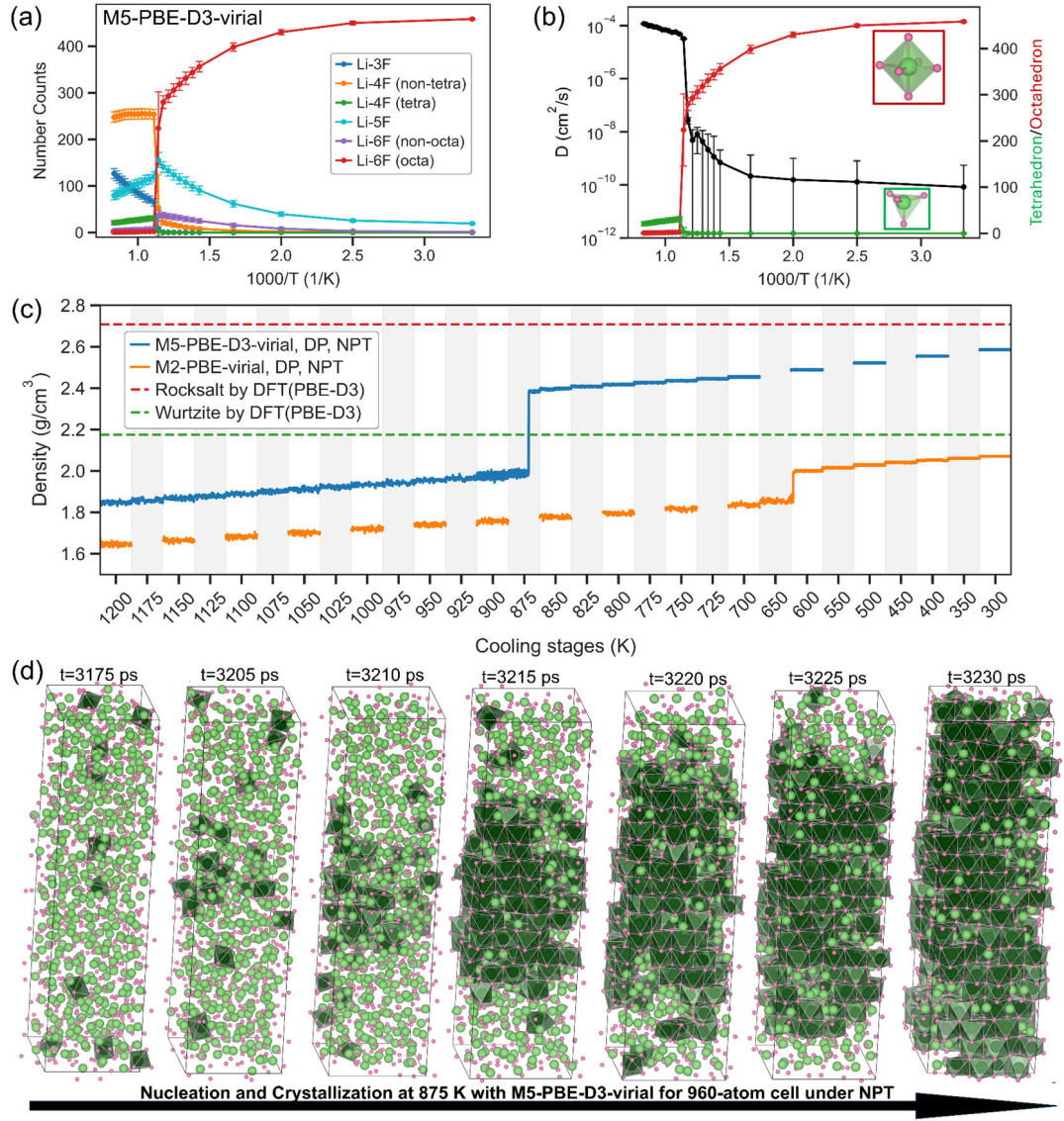

**Figure S10.** Model trained with our own dataset with PBE-D3 from AIMD simulations and random perturbations of ordered wurtzite and rocksalt LiF supercells. (a) Coordination statistics for 960-atom LiF supercell from MLMD trajectories. (b) Li diffusion coefficients (black), the octahedrons (red) and tetrahedrons (green) numbers in the LiF as functions of temperatures. (c) Comparisons of density evolution through NPT simulations with **M2-PBE-virial** and **M5-PBE-D3-virial**. The PBE-D3-calculated wurtzite and rocksalt LiF densities were shown as dashed green and red lines. (d) The nucleation and crystallization process captured by **M5-PBE-D3-virial** NPT simulation at 875 K. The Li-6F units were highlighted in dark green.

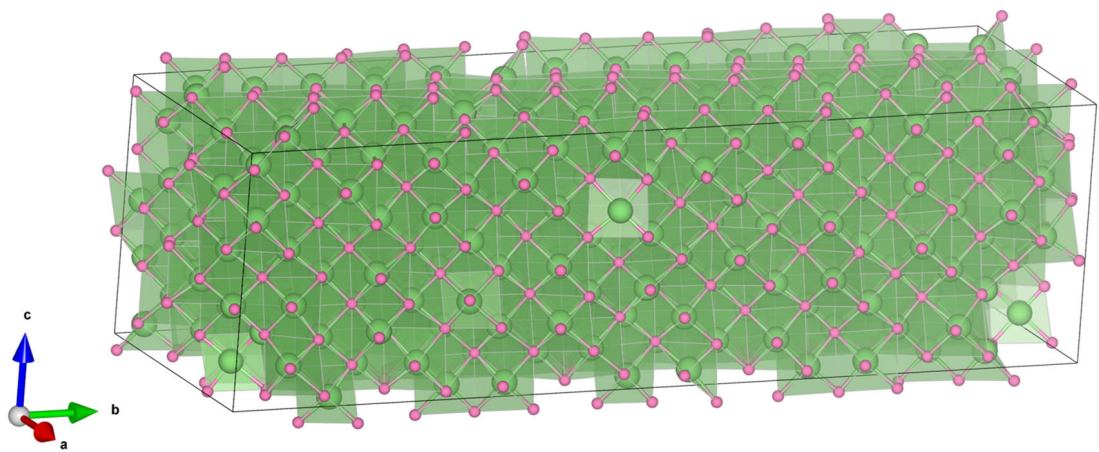

**Figure S11.** A snapshot of a specific M5 structure after phase change at 875 K, showing the existence of some vacuum space (0.12%) at right bottom corner.

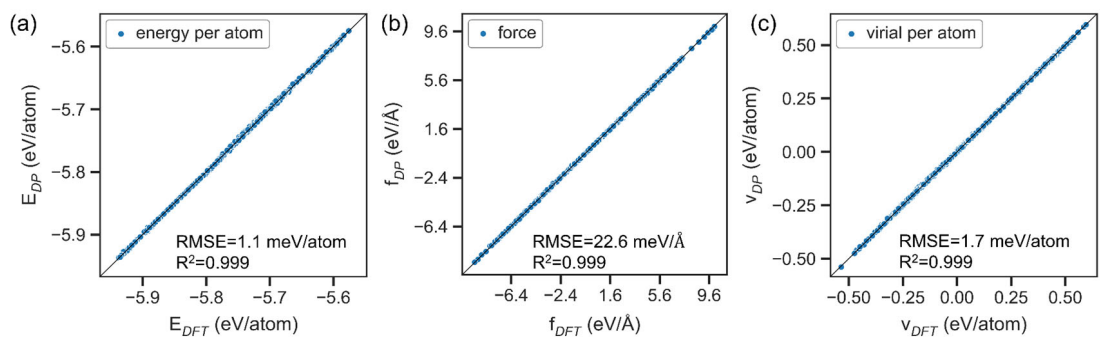

**Figure S12.** Comparisons of model trained with $r^2$SCAN method: (a) energy comparison for DP *vs* DFT, (b) force comparison for DP *vs* DFT and (c) virial per atom for DP *vs* DFT.

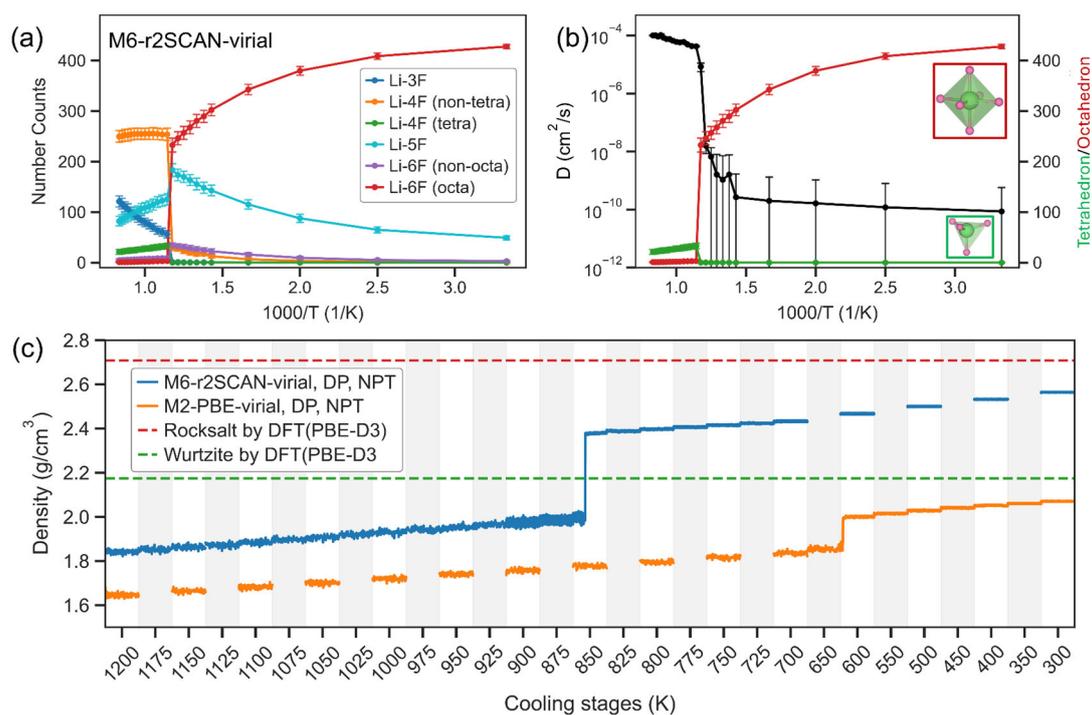

**Figure S13.** Model trained with our own dataset with $r^2$SCAN from AIMD simulations and random perturbations of ordered wurtzite and rocksalt LiF supercells. (a) Coordination statistics for 960-atom LiF supercell from MLMD trajectories. (b) Li diffusion coefficients (black), the octahedrons (red) and tetrahedrons (green) numbers in the LiF as functions of temperatures. (c) Comparisons of density evolution through NPT simulations with **M2-PBE-virial** and **M6-$r^2$SCAN-virial**. The PBE-D3-calculated wurtzite and rocksalt LiF densities were shown as dashed green and red lines.

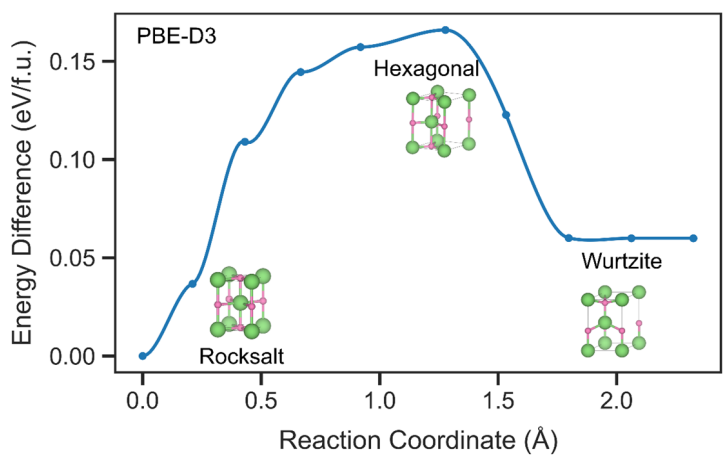

**Figure S14.** The SSNEB results showed that the lowest energy evolution path for rocksalt to wurtzite transition went through hexagonal intermediates.

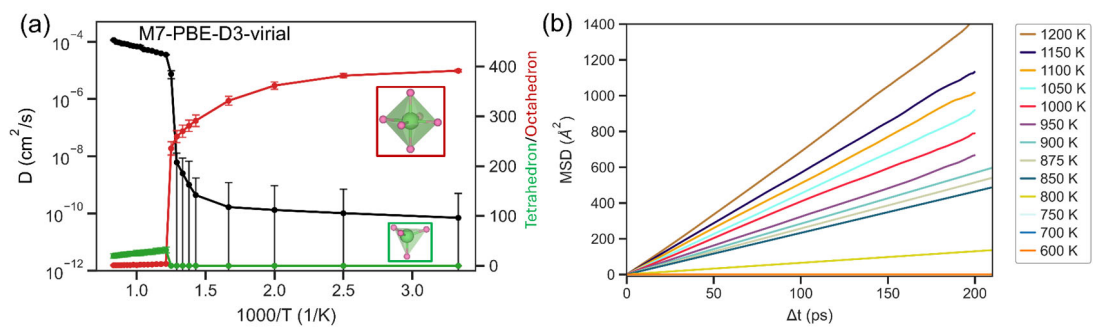

**Figure S15.** (a) The analyzed Li diffusivities and (b) first 200 ps MSD from 960-atom **M7-PBE-D3-virial** NPT simulations. The defined octahedron (red) and tetrahedron (green) numbers in the LiF as a function of temperatures were plotted。

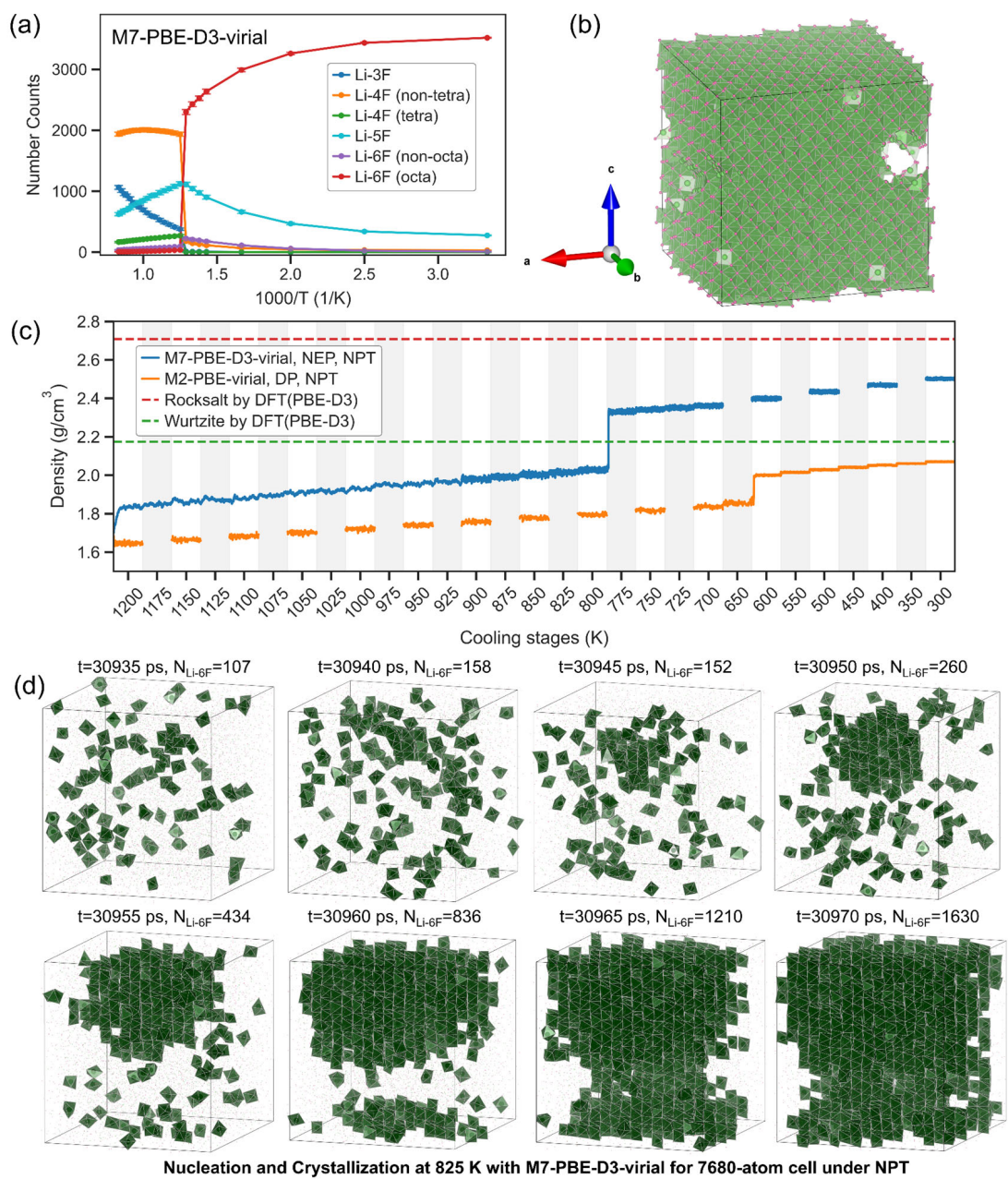

**Figure S16.** NEP-ZBL model trained with our own dataset with PBE-D3 from AIMD simulations and random perturbations of ordered wurtzite and rocksalt LiF supercells. (a) Coordination statistics for 7680-atom LiF supercell from MLMD trajectories. (b) Defective structure with obvious vacuum space (1.41%) at the last frame of 300 K was plotted. (c) Comparisons of density evolution through NPT simulation with **M2-PBE-virial** and **M7-PBE-D3-virial**. The PBE-D3-calculated wurtzite and rocksalt LiF densities were shown as dashed green and red lines. (d) The nucleation and crystallization process captured by **M7-PBE-D3-virial** NPT simulation at 825 K (the long run). The Li-6F units were highlighted in dark green and their numbers were counted and listed.
14

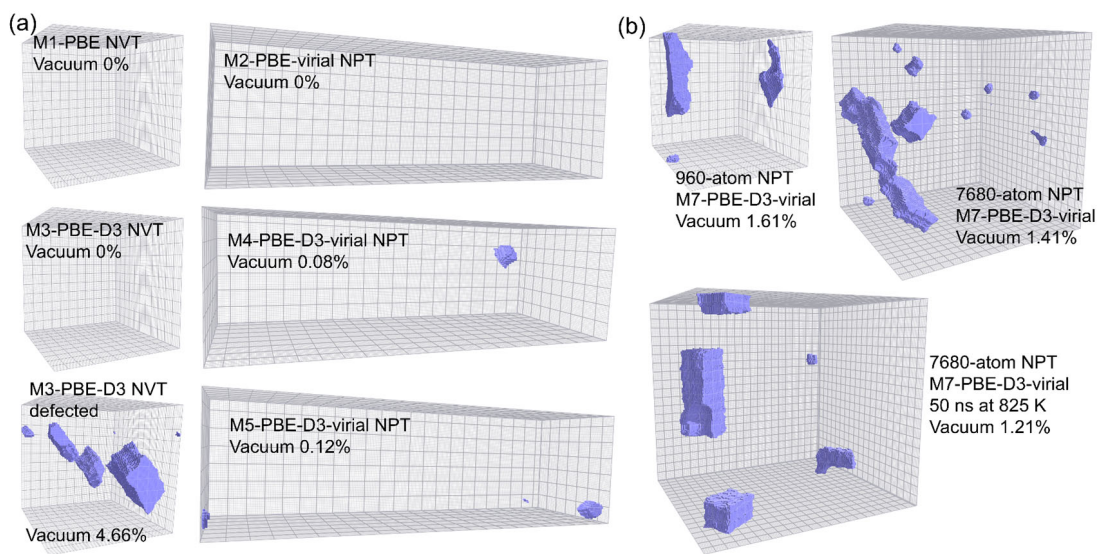

**Figure S17.** Vacuum analysis visualized result for (a) the last frame of 300 K structures generated by MLMD with DP (b) the last frame of 300 K 960-atom and 7680-atom MLMD structures generated by MLMD with NEP. Besides, the last frame structure of separated 50 ns 7680-atom simulation at 825 K was also included. The ratios of vacuum volume over cell volume were reported.

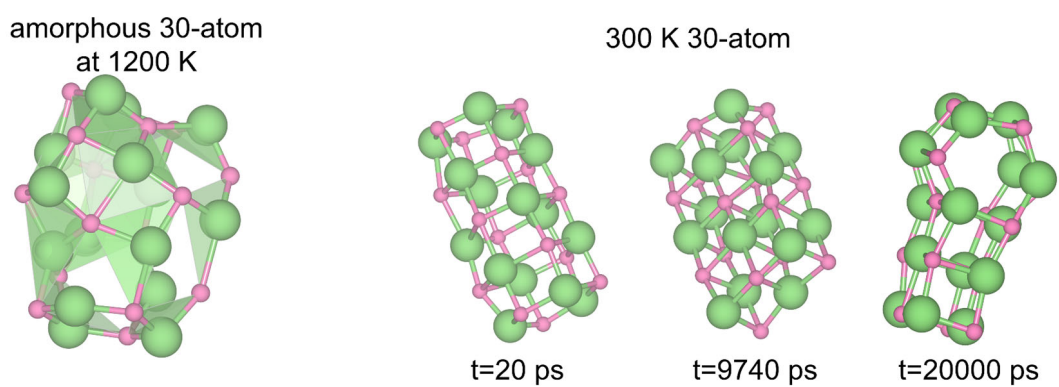

**Figure S18.** The 30-atom cluster did not present a determined geometry even at 300 K, the structures at 20 ps, 9740 ps and 20000 ps were picked and showed different configurations.